%% file: main.tex
\documentclass[sigconf]{acmart}

\usepackage{booktabs} 
\usepackage{subfigure}
\usepackage{color}
\usepackage{tabu}
\usepackage{physics}
\usepackage{epigraph}
\usepackage{ctable}
\usepackage{enumitem}
\setlist[itemize]{leftmargin=*}

\copyrightyear{2020}
\acmYear{2020}
\setcopyright{acmcopyright}
\acmConference[WSDM '20]{The Thirteenth ACM International Conference on Web Search and Data Mining}{February 3--7, 2020}{Houston, TX, USA}
\acmBooktitle{The Thirteenth ACM International Conference on Web Search and Data Mining (WSDM '20), February 3--7, 2020, Houston, TX, USA}
\acmPrice{15.00}
\acmDOI{10.1145/3336191.3371776}
\acmISBN{978-1-4503-6822-3/20/02}

\begin{document}
\fancyhead{}

\title{Consistency-Aware Recommendation for User-Generated Item List Continuation}


\author{Yun He}
\affiliation{%
  \institution{Texas A\&M University}
}
\email{yunhe@tamu.edu}

\author{Yin Zhang}
\affiliation{%
  \institution{Texas A\&M University}
}
\email{zhan13679@tamu.edu}

\author{Weiwen Liu}
\affiliation{%
  \institution{The Chinese University of Hong Kong}
}
\email{wwliu@cse.cuhk.edu.hk}

\author{James Caverlee}
\affiliation{%
  \institution{Texas A\&M University}
}
\email{caverlee@tamu.edu}

\begin{abstract}
User-generated item lists are popular on many platforms. Examples include video-based playlists on YouTube, image-based lists (or ``boards'') on Pinterest, book-based lists on Goodreads, and answer-based lists on question-answer forums like Zhihu. As users create these lists, a common challenge is in identifying what items to curate next. Some lists are organized around particular genres or topics, while others are seemingly incoherent, reflecting individual preferences for what items belong together. Furthermore, this heterogeneity in item consistency may vary from platform to platform, and from sub-community to sub-community. Hence, this paper proposes a generalizable approach for user-generated item list continuation. Complementary to methods that exploit specific content patterns (e.g., as in song-based playlists that rely on audio features), the proposed approach models the consistency of item lists based on human curation patterns, and so can be deployed across a wide range of varying item types (e.g., videos, images, books). A key contribution is in intelligently combining two preference models via a novel consistency-aware gating network -- a general user preference model that captures a user's overall interests, and a current preference priority model that captures a user's current (as of the most recent item) interests. In this way, the proposed consistency-aware recommender can dynamically adapt as user preferences evolve. Evaluation over four datasets (of songs, books, and answers) confirms these observations and demonstrates the effectiveness of the proposed model versus state-of-the-art alternatives. Further, all code and data are available at https://github.com/heyunh2015/ListContinuation\_WSDM2020.
\end{abstract}

%
%
%



\maketitle

\input{samplebody-conf}

\begin{acks}
This work is supported in part by NSF (\#IIS-$1841138$). 
\end{acks}

\bibliographystyle{ACM-Reference-Format}
\bibliography{sample-bibliography}

\end{document}

%% file: samplebody-conf.tex
\section{Introduction}
\label{introduction}
%

\begin{figure}
  \centering
   \setlength{\abovecaptionskip}{0.01cm}
  \setlength{\belowcaptionskip}{-0.77cm}
 \subfigure[List title: ``Fantasy books"]{
    \label{list with Strong consistency}
    \includegraphics[width=1.55in]{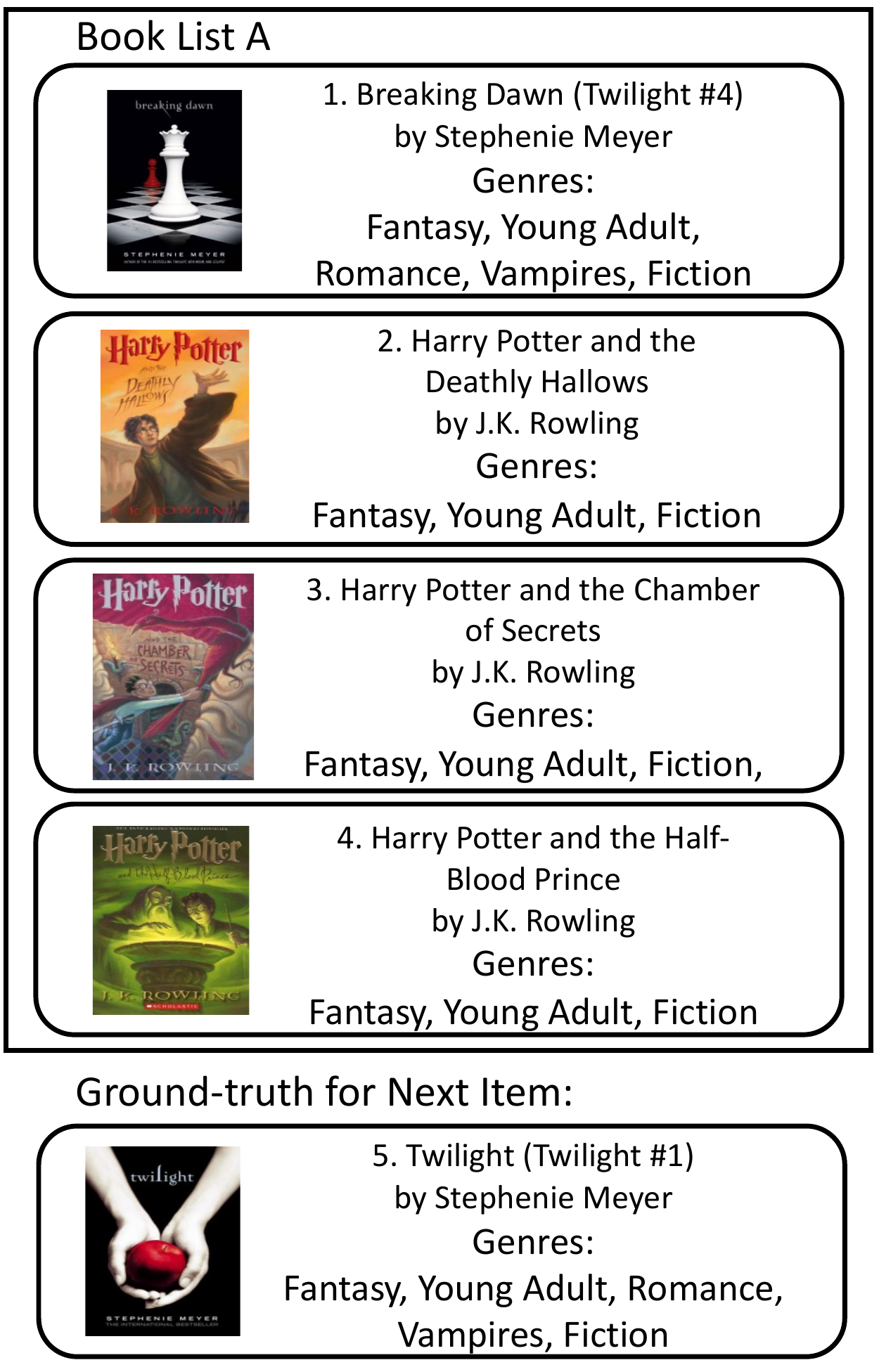}}
 \subfigure[List title: ``My favorite books in 2019"]{
    \label{A list with Weak consistency} 
    \includegraphics[width=1.55in]{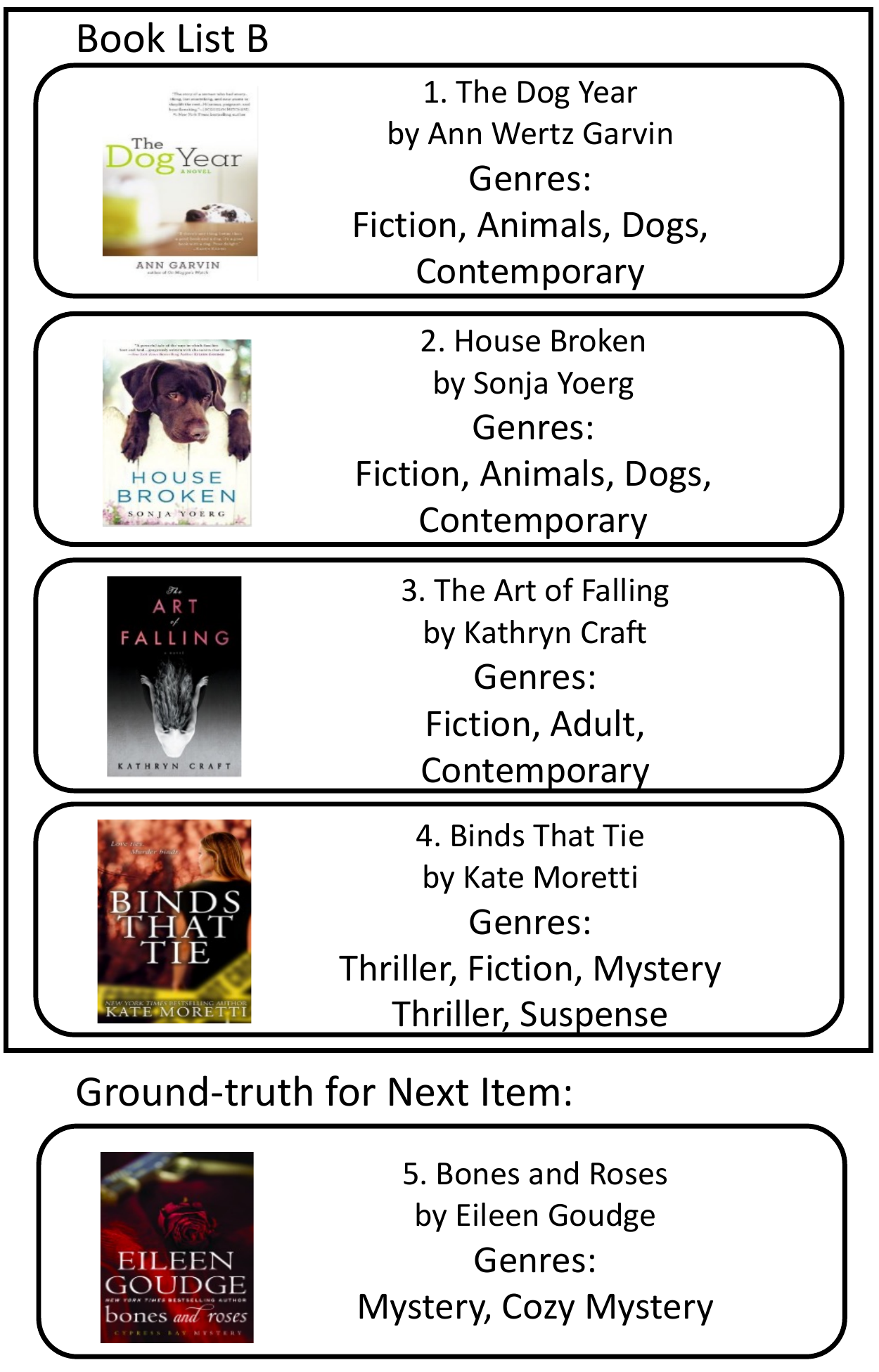}}
   \caption{A motivating example of two lists with strong and weak consistency between recent items and previous items. }
  \label{Motivating Example} 
\end{figure}

Human curation is a widely used feature in platforms like Spotify, Pinterest, YouTube, and Goodreads. Users can curate items like songs, images, videos and books to form lists that provide a unique perspective into how items can be grouped together. For example, Figure~\ref{Motivating Example} shows two book lists on the book sharing platform Goodreads; one is organized around a genre (fantasy) while the other collects a personal list of favorites spanning genres. Since correlated items can be explored and consumed together, these item lists directly power user engagement. For example, more than 50\% of Spotify users listen to playlists, accounting for more than 1 billion plays per week \cite{buzzfeed2016}; and Pinterest users have curated more than 3 billion pins to about 4 billion boards \cite{eksombatchai2018pixie}.

In these platforms, user-generated item lists are manually created, curated, and managed by users. Typically, users must first identify candidate items, determine if they are a good fit for a list, add them to a list, and then potentially provide ongoing updates to the list (e.g., by adding or deleting items over time). To accelerate this process and assist users to explore more related items for their lists, we study the important yet challenging problem of \textit{user-generated item list continuation}. That is, how can we recommend items that are related to the list and fit the user's preferences? Compared with traditional item-based recommendation \cite{bpr}, list continuation faces complexities, since the length of lists varies from several items to even thousands of items and the user preference on items may dynamically evolve as a list develops. Moreover, the individual preference of how items are grouped together also varies for each user and across different platforms. Facing these challenges, we propose a consistency-aware recommender for user-generated item list continuation that is motivated by three key observations.

\begin{figure*}
  \centering
\setlength{\abovecaptionskip}{0.0cm}
  \setlength{\belowcaptionskip}{-0.5cm}
 \subfigure[Song-based lists in Art of the Mix]{
    \label{AOTM consistency}
    \includegraphics[width=1.6in]{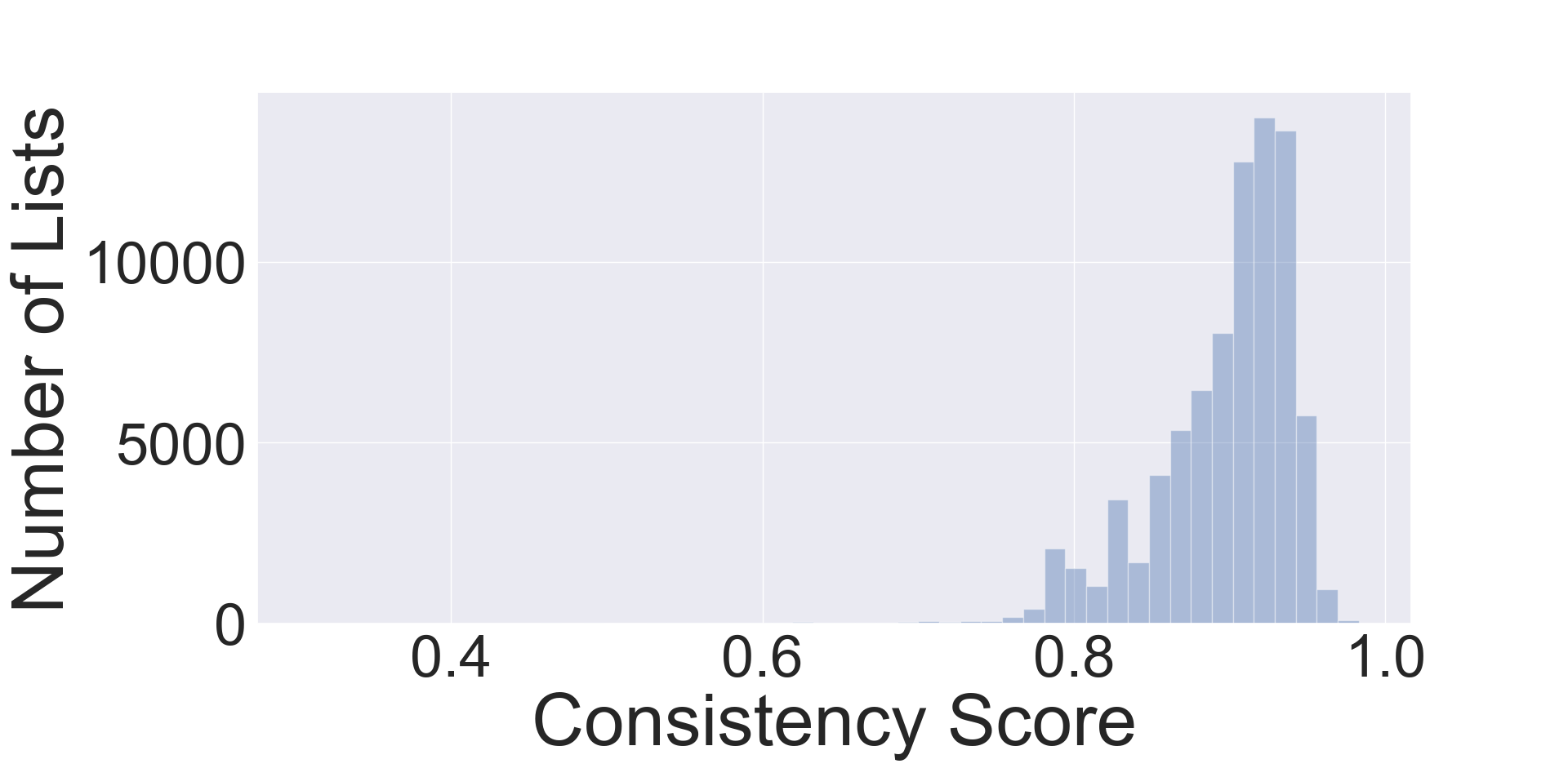}}
   \subfigure[Song-based lists in Spotify]{
    \label{Spotify consistency} 
    \includegraphics[width=1.6in]{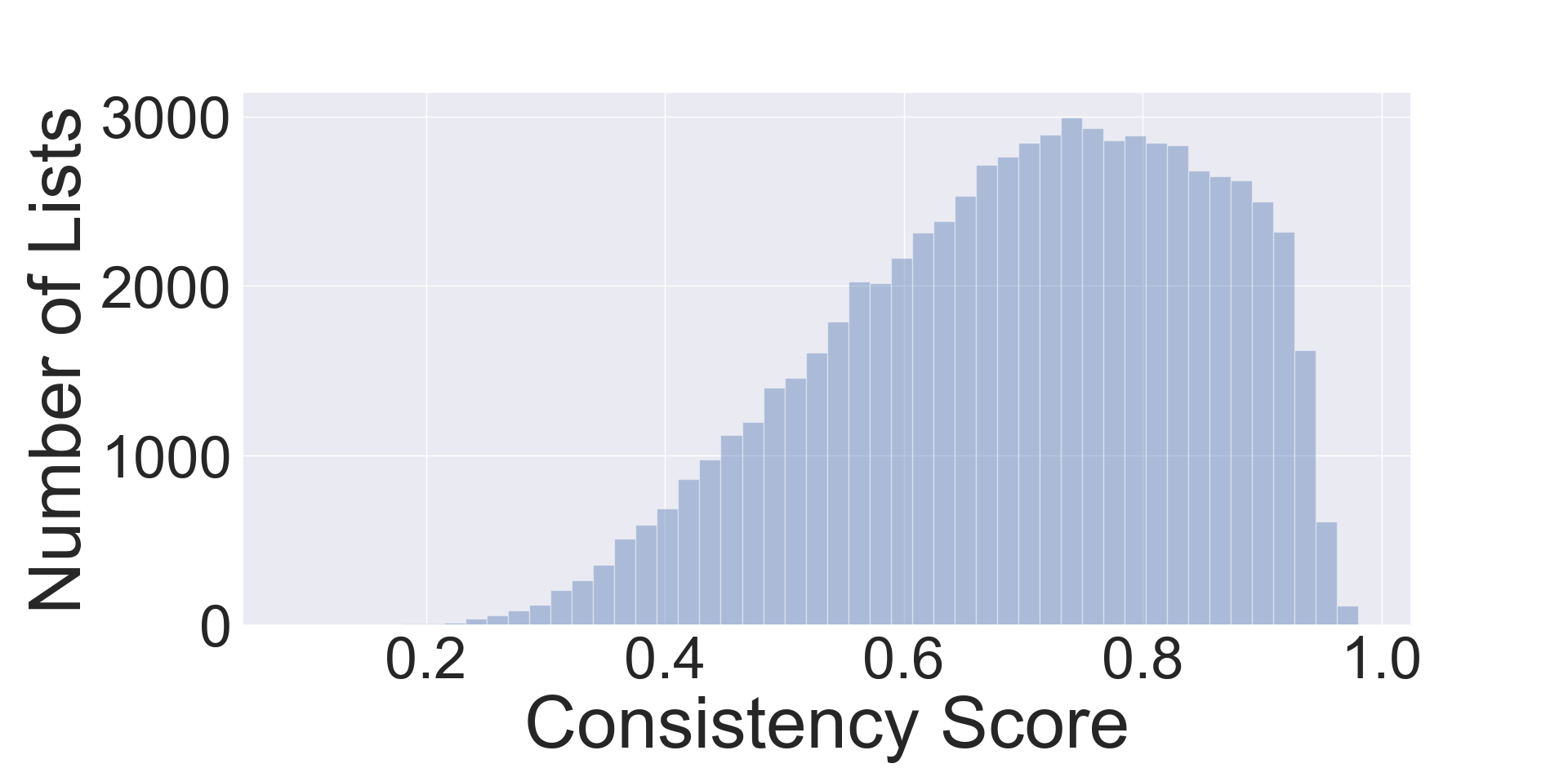}}
 \subfigure[Book-based lists in Goodreads]{
    \label{Goodreads consistency} 
    \includegraphics[width=1.6in]{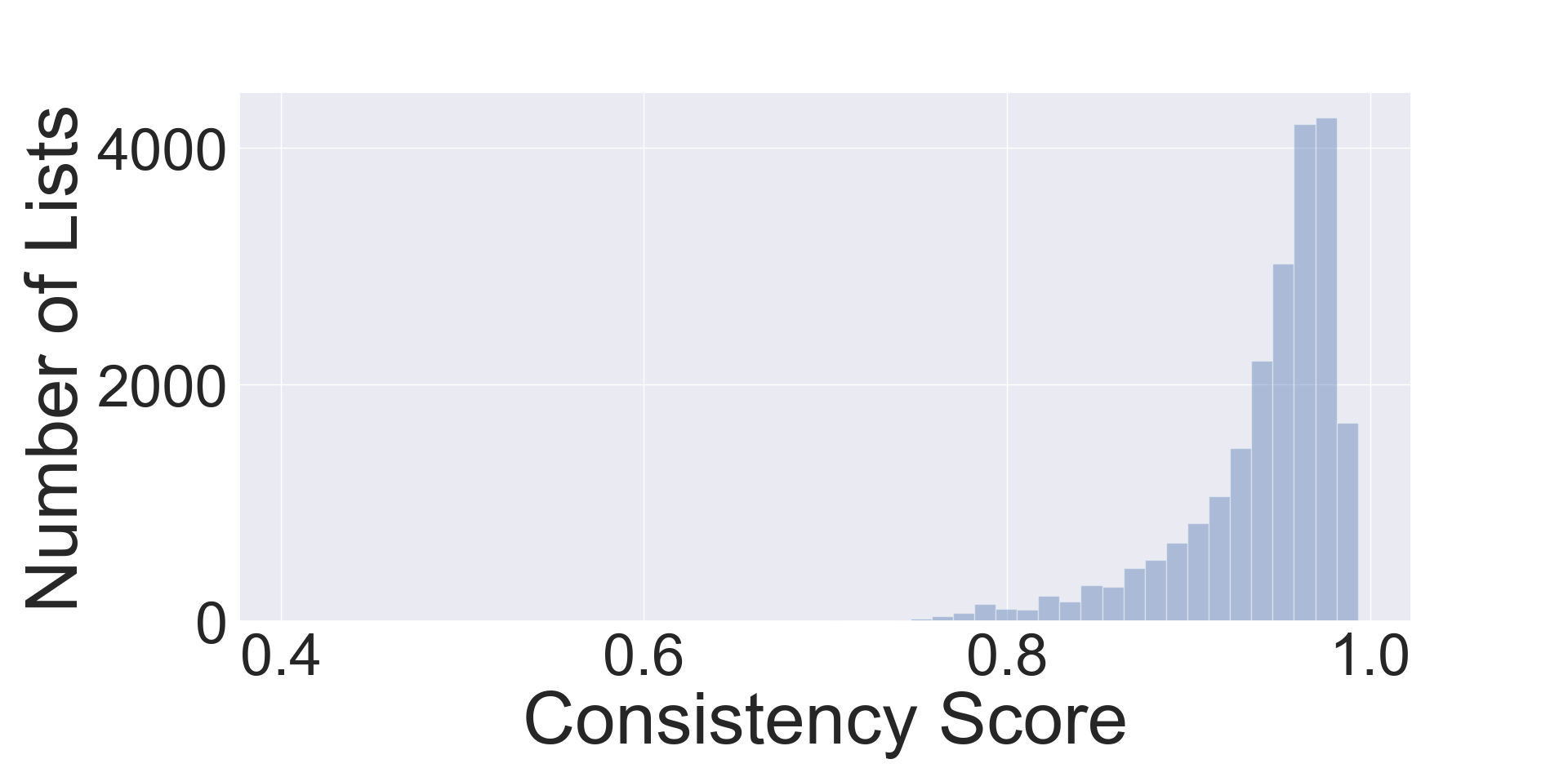}}
  \subfigure[Answer-based lists in Zhihu]{
    \label{Zhihu consistency} 
    \includegraphics[width=1.6in]{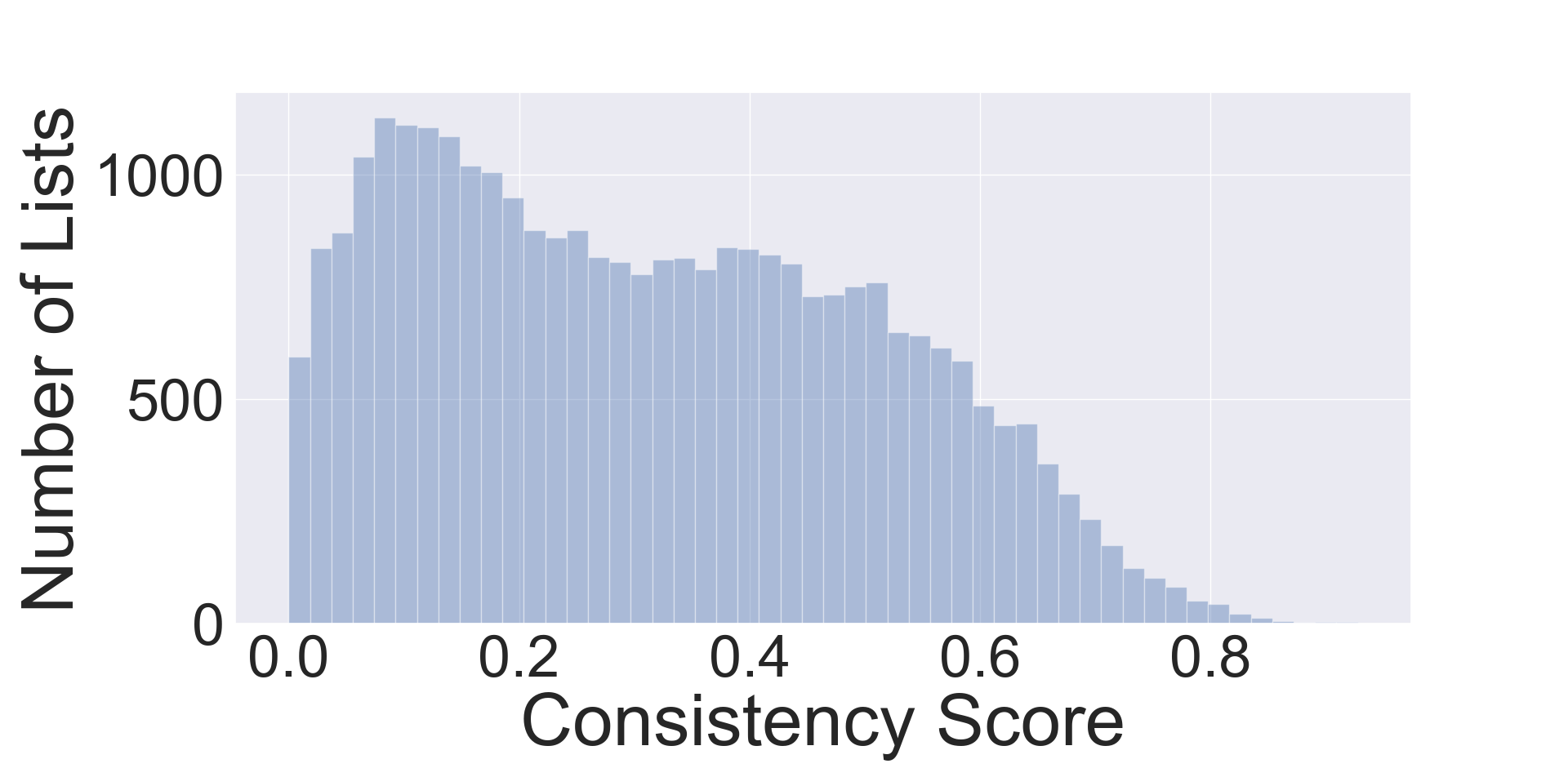}}
   \caption{The consistency between recent items and previous items (by Eq.\eqref{consistency equation}). We observe that the list consistency varies greatly both within a single platform and across different platforms.}
  \label{histogram of consistency} 
\end{figure*}

First, we observe that \textit{some lists are strongly consistent while others are only weakly consistent.} Traditional song-based playlist continuation usually aims to generate strongly consistent lists, where each new song naturally fits with the previous ones \cite{platt2002learning, bonnin2015automated, chen2012playlist, hariri2012context, jannach2015beyond, mcfee2011natural, mcfee2012hypergraph, flexer2008playlist, maillet2009steerable, chen2018recsys, tran2019adversarial}. A typical example is in \cite{platt2002learning}, which generates a playlist with songs that are similar to the seed songs provided by the user. Indeed, some user-generated item lists may be organized around a specific theme, genre, or mood. However, some lists are organized by more personalized patterns and are seemingly inconsistent. To illustrate, List A in Figure~\ref{list with Strong consistency} has strong consistency between recent books and previous books. The books are from either the \textit{Harry Potter} series or the \textit{Twilight} series, organized around genres like ``Fantasy". List B in Figure \ref{A list with Weak consistency} is only weakly consistent with a mystery book \textit{Binds That Tie} following two books about dogs. Indeed, we find in  Figure~\ref{histogram of consistency} that list consistency varies greatly both within a single platform and across different platforms. As a result, we require new methods that can dynamically adapt to this heterogeneity of list consistency. 




Second, we observe that \textit{list consistency has a strong impact on the quality of predicting the next item.} The examples in Figure \ref{Motivating Example} show that the ground-truth for strongly consistent list A is the fantasy book in the \textit{Twilight} series while the ground-truth for the weakly consistent list B is a mystery book that is similar with the most recent books in the list, but not the books earlier in the list. Our intuition is that if recent items are not similar with previous items, the user preference has probably changed recently. Therefore, the recent items (reflecting the current user preference) should be given more attention when we predict the next possible item. On the contrary, if recent items are still consistent with previous ones, the user preference is probably unchanged and so the general user preference should pay equal attention to items no matter if they have been curated early or late.


Third, we observe that \textit{the definition of what makes a list consistent should arise from community norms rather than be imposed top-down}. Prior work in song-based playlists has found that neither audio signal-based similarity nor social tag-based similarity accurately reflect the consistency of manually constructed playlists \cite{chen2012playlist, mcfee2011natural}. Our example so far in Figure~\ref{Motivating Example} has highlighted how books in a single list may be drawn from different (seemingly inconsistent) genres. However, the consistency of items should depend on how users in the community perceive those items; so, if many users curate two items together, then those items should be considered consistent regardless of their superficial similarity. Hence, item list continuation methods should seek to model consistency based on the curation patterns arising from the community itself.




These observations motivate us to attack the user-generated item list continuation problem with a \textbf{C}onsistency-aware and \textbf{A}ttention-based \textbf{R}ecommender (CAR). First, we propose both a general user preference model that aims to capture a user's overall interests for the case of strongly consistent lists, and a current preference priority model that captures a user's current interests for cases where consistency adapts over time. Second, to provide a generalizable approach that can handle the heterogeneity in list consistency across lists in different platforms, we design a novel consistency-aware gating network to adaptively balance and fuse the current and general user preferences, where the list consistency is modeled by computing the variation of the items in a list.

We evaluate CAR over four datasets reflecting different kinds of item lists and different curation patterns: the song-based Art of the Mix (AOTM) and Spotify, book-based  Goodreads, and the answer-based Zhihu platform. Through experiments versus a suite of state-of-the-art baselines, we find that CAR significantly outperforms them and our general user preference model and current preference priority model can really complement each other guided by list consistency. Further, all code and data will be released to the research community for further exploration.

\section{Related Work}
\noindent\textbf{Automatic Playlist Generation and Continuation.} Considerable prior research has focused on automatic song-based playlist generation and continuation, aligned along three branches. The first branch focuses on generative models for estimating the likelihood of a new playlist by treating training playlists as a set of song sequences. For example, McFee et al. \cite{mcfee2012hypergraph, mcfee2011natural} apply a first-order Markov Chain for modeling playlists, which is improved by Chen et al. \cite{chen2012playlist} via introducing metric embeddings. The second branch mainly relies on song-based audio features (e.g., Mel-frequency cepstral coefficients) to generate new playlists. Examples include \cite{flexer2008playlist} and \cite{maillet2009steerable}, where Maillet et al. \cite{maillet2009steerable} train classifiers using audio-based features to determine if a sequence of songs can form a playlist. The third branch is most similar to our work, often applying information retrieval or recommendation methods to predict (or continue) the next $N$ items. These approaches often are enhanced by content-based methods \cite{platt2002learning, hariri2012context, jannach2015beyond, zamani2018analysis, volkovs2018two, zhao2018trailmix}. A typical example by Jannach et al. \cite{jannach2015beyond} uses collaborative filtering and incorporates content features like social tags from Last.fm and popularity of songs.

\medskip
\noindent\textbf{Differences:} There are two key differences between our work and traditional song-based playlist generation. First, we focus on not only song-based lists but propose a general framework that can be applied to other user-generated item lists, like book-based (Goodreads book lists), video-based (YouTube playlists), image-based (Pinterest boards) and answer-based (Zhihu collections) lists. Like song-based playlists, each collection of these item lists also provides a unique resource where correlated items can be easily explored and consumed together, directly empowering user engagement. Therefore, automatic generation and continuation for these item lists should be equally important as song-based playlists. Second, most traditional playlist continuation work assumes lists are always consistent \cite{platt2002learning, chen2012playlist, hariri2012context, jannach2015beyond, mcfee2011natural, mcfee2012hypergraph, flexer2008playlist, maillet2009steerable}. Except for \cite{platt2002learning} introduced in the introduction, examples include \cite{mcfee2012hypergraph, mcfee2011natural, chen2012playlist} that apply language models to generate coherent playlists and \cite{jannach2015beyond} that recommends a set of songs whose tempo distribution is as similar as possible to the current playlist. However, we observe that the consistency varies greatly for lists both within and across different platforms, meaning that list continuation methods should dynamically adapt to these scenarios.

\medskip
\noindent\textbf{User-Generated Item Lists.} Recently, user-generated item lists have received more and more interest. Zhong et al. \cite{zhong2013sharing} study the motivations of human curation to reveal the social values of user-generated item lists. Feinberg et al. \cite{feinberg2012understanding} study the characteristics of item lists and claim there are two kinds of lists: for personal information management and for public expression. Lo et al. \cite{lo2017understanding} analyze the growth of image collections on Pinterest. Lu et al. \cite{liu2016power} and Eksombatchai et al. \cite{eksombatchai2018pixie} distill user preference from Pinterest image-based lists to enhance individual image recommendation. Another interesting problem is to recommend existing item lists to users. He et al. \cite{he2019hierarchical} propose a hierarchical self-attentive model for recommending user-generated item lists (e.g., book lists and playlists) to right users. Besides, the List Recommending Model in \cite{liu2014recommending} is proposed for recommending book lists and the Embedding Factorization Model in \cite{cao2017embedding} is for recommending song playlists. 

\medskip
\noindent\textbf{Neural Networks for Recommendation.} Recently, neural networks have been widely applied in recommendation. He et al. \cite{ncf} apply multilayer perceptron and generalized matrix factorization for implicit top-k recommendation. Ebesu et al. \cite{ebesu2018collaborative} apply memory networks for recommendation. Hidasi et al. \cite{hidasi2015session} propose a GRU-based model for session-based recommendation and Tang et al. \cite{tang2018personalized} propose a CNN-based model for sequential recommendation. Attention networks are often used for weighted-summing elements in a model. \cite{chen2017attentive} propose an attentive collaborative filtering framework, where each item is segmented into component-level elements, and attention scores are learned for these components for obtaining a better representation of items. Attention networks are also applied in group recommendation \cite{cao2018attentive} and sequential recommendation \cite{kang2018self, huang2018csan}. 

\section{Preliminaries}
\label{Preliminaries}
In this section, we formulate the problem of user-generated item list continuation and present data-driven evidence to motivate our approach.

\subsection{User-generated Item List Continuation} 
\label{Problem Formulation}
Let $U = \{u_{1}, u_{2},..., u_{|U|}\}$ denote a set of users and a set of unique items is denoted by $V=\{v_{1}, v_{2},..., v_{|V|}\}$. Each list generated by a user $u$, denoted by $S^{u}=[s_{1}, s_{2},..., s_{N}]$, consists of a sequence of $N$ items from set $V$. Moreover, $S^{u}_{t} = [s_{1}, s_{2}, ..., s_{t}]$, $1 \leqslant t \leqslant N$ denotes a truncated item list at item-step $t$ with regard to list $S^{u}$, where $s_{i}$ is an item curated at time-step $i$. The goal of user-generated item list continuation is to predict the next possible curated item (i.e., $s_{t+1}$) based on the given $S^{u}_{t}$ (for simplicity, we use $S$ to denote $S^{u}_{t}$ in the rest of the paper). Specifically, the recommender outputs a score for each item in the candidate set $D$ ($D\in V$). After that, items in $D$ can be ranked by their scores in descending order and top items can be returned for the item list continuation. 

\subsection{Evidence of Strong and Weak Consistency}
\label{motivation}
The two examples in Figure \ref{Motivating Example} highlight a strongly consistent list and a weakly consistent list. Here, we present a brief data-driven examination to further explore such consistency across four different platforms: AOTM, Spotify, Goodreads and Zhihu (details of these datasets can be found in Section \ref{datasets}).

\medskip
\noindent\textbf{Measuring Consistency.} To measure consistency of a user-generated item list $S$, we propose to measure the average of similarities between recent items and each of the previous items. So a strongly consistent list will have a high average similarity of new items that are added compared to the previous items. A weakly consistent list will have a low average similarity as new items begin to deviate from previous items. In contrast to related approaches like Intra-List Similarity proposed in \cite{ziegler2005improving} that measure the average of pairwise similarities between all items in a list, this consistency measure is designed to capture how a human curator may incrementally add items to the list. For simplicity, if we use the last item to represent recent items, then the consistency is:

\begin{equation}
\label{consistency equation}
	consistency\text{-}score_{S} = \sum_{s_{i} \in S, i \neq t} \frac{sim(s_{i}, s_{t})}{N} 
\end{equation}
where $N$ is the length of a list, $s_{t}$ is the last item of list $S$ and $sim(s_{i}, s_{t})$ is the similarity between $s_{i}$ and $s_{t}$. Of course there are many ways to measure this item-based similarity including collaborative filtering \cite{sarwar2001item}, co-occurrence of items \cite{mikolov2013distributed} and content-based similarity. For this analysis we adopt a variation of word2vec \cite{mikolov2013distributed} where items are ``words'' and lists are ``sentences''. This approach captures the co-occurrence of items to generate item embeddings, which can then be used to calculate the cosine similarity between the embeddings: $sim(s_{i}, s_{t}) = cosine(e_{i}, e_{t})$, where $e_{i}$, $e_{t}$ are the embeddings of $s_{i}$, $s_{t}$ derived from word2vec. Note that our proposed approach in Section~\ref{Fusing Long and Short-Term Preferences} adopts a recommendation model-driven representation for items and lists instead of this word2vec-style approach.

\medskip
\noindent\textbf{Observations:} For each of our four datasets -- AOTM, Spotify, Goodreads and Zhihu -- the histogram of $consistency\text{-}score_{S}$ is presented in Figure \ref{histogram of consistency}. We observe that \textit{the list consistency varies greatly both within a single platform and across different platforms.} For example, Spotify lists cover a wide range of consistency values, with some lists having a $consistency\text{-}score_{S}$ more than 0.8 while others are lower than 0.4. Both AOTM and Goodreads have scores that tend to skew high, while the answer-based lists on Zhihu tend to be lower. Furthermore, this heterogeneity of list consistency could negatively impact the quality of predicting the next item. For example, a traditional playlist recommender that expects all items to maintain high consistency will suggest items that do not match the current user preference as suggested by the examples in Figure \ref{Motivating Example} and the data-driven evidence here. Therefore, we are motivated to propose a consistency-aware recommendation in the next section.

\section{Proposed Approach: CAR}
\label{Fusing Long and Short-Term Preferences}

Motivated by this heterogeneity of list consistency, we propose a general user preference model to capture the user's overall interests for strongly consistent lists, and a current preference priority model to emphasize more on the user's current interests for lists with weak consistency. Moreover, as list consistency varies across lists in different platforms, we design a novel consistency-aware gating network to adaptively combine the two models. As discussed in Section \ref{motivation}, we propose to measure the consistency by computing the variation of the items in a list to aware if the current preference has deviated from the general user preference. 

\subsection{Attention-based User Preference Model}
\label{Attention-based User Preference Model}

To predict the next possible item for each list, the key problem is to model the user preference from the items already curated in the list. A natural way is to first aggregate the item embeddings to represent the list embedding, and then match the obtained list embedding with other item embeddings to predict the next possible item. 

To adaptively aggregate the item embeddings in the list, we design two aggregation schemes with attention networks -- the general user preference model and the current preference priority model. Attention networks have been widely applied in recommendation to weighted-sum a variety of components (e.g., items \cite{he2018nais, chen2017attentive} and user behaviors \cite{zhou2018micro, zhou2017atrank}), achieving promising performance. Following the terminologies in \cite{vaswani2017attention}, attention mechanism can be formalized as:
\begin{equation}\label{attention}
\begin{gathered}
  \alpha_{i} = softmax(k_{i}^{T}q_{i}), \\
  Attention(\{v_{i}, k_{i}, q_{i}|i=1,...,t\}) = \sum_{i=1}^{t} \alpha_{i}v_{i}
\end{gathered}
\end{equation}

\noindent where we first compute the inner product ($k_{i}^{T}q_{i}$) of ($k_{i}$) and query ($q_{i}$), then the weight ($\alpha_{i}$), which reflects the importance of values ($v_{i}$), is obtained by normalizing over all key-query pairs with a softmax function ($softmax(z_{i})=\frac{e^{z_{i}}}{\sum_{j=1}^{t}e^{z_{j}}}$). The result is a weighted-sum by aggregating values ($v_i$). More clearly:

\begin{itemize}
	\item Values ($v_{i}$) are components await to be aggregated together.
	\item Keys ($k_{i}$) correspond to values that are normally linear or non-linear transformed from values.
	\item Queries ($q_{i}$) play a role of benchmarking to measure the importance of values by matching against keys.
\end{itemize}

\medskip
\noindent\textbf{Embedding Layer:} Let $Y=\{y_{1}, y_{2},...,y_{|U|}\}$ denote the embedding vectors of users in $U$, where $y_{u}\in \mathbb{R}^{d}$ is a learnable d-dimensional real-valued vector for user $u$ in $U$. Likewise, let $X=\{x_{1}, x_{2},...,x_{|V|}\}$ denote the embedding vectors of items in set $V$, where $x_{i}\in \mathbb{R}^{d}$ is the i-th item embedding in list $S$. Note that $x_{t}\in \mathbb{R}^{d}$ represents the embedding of the last (most recent) item curated in the list at time-step $t$ (i.e., $s_{t}$).

\medskip
\noindent\textbf{Design of Values and Keys:} For both preference models, we first let $v_{i} = x_{i}$ in Eq.\eqref{attention}, since the goal of a user preference model is to aggregate item embeddings to represent a list embedding. After that, we let $k_{i} = W^{K}x_{i}$, where $W^{K}\in \mathbb{R}^{d \times d}$ is a linear projection to map $x_{i}$ into the ``keys" space for matching against queries. The critical difference between the general and current preference priority model is how to choose queries ($q_i$), elaborated as follows. 

\medskip
\noindent\textbf{General User Preference Model (GUPM):} In strongly consistent lists, the user preference is probably unchanged. Hence, we design GUPM to model the overall user preferences in the list by treating items equally important, no matter at which time step the items have been curated. A natural way is to incorporate another learnable vector $h \in \mathbb{R}^{d}$ as the query, normally referred to as global or context vector in the literature \cite{cao2018attentive, chen2017attentive}. We have:

\begin{equation}
\label{vanilla attention equation}
\begin{gathered}
  \alpha_{i} = softmax((W^{K}x_{i})^{T}h), \\
 l_{S}^{G} = \sum_{i=1}^{t} \alpha_{i}x_{i}
\end{gathered}
\end{equation}

\noindent where $\forall \ i \ \in \ \{1,...,t\}, q_{i}=h$ denotes the query and $l_{S}^{G}$ is the list embedding generated by the GUPM.

\medskip
\noindent\textbf{Current Preference Priority Model (CPPM):} In weakly consistent lists, recent items begin to deviate from previous items and should be given more attention for predicting the next item. CPPM captures user's current interests by assigning higher priority on recent item over other items. A natural way is to let the last item as the benchmark to measure the importance of other items: the more similar to the last item the higher weight is assigned. Hence, we let the last item embedding be the query:

\begin{equation}
\label{self attention equation}
\begin{gathered}
  \alpha_{i} = softmax((W^{K}x_{i})^{T}W^{Q}x_{t}), \\
  l_{S}^{C} = \sum_{i=1}^{t} \alpha_{i}x_{i}
\end{gathered}
\end{equation}

\noindent where $x_{t}$ is the embedding of the last item, $\forall \ i \ \in \ \{1,...,t\}, q_{i}=W^{Q}x_{t}$ denotes the query, $W^{Q}\in \mathbb{R}^{d \times d}$ is a linear projection to map $x_{t}$ into ``queries" space for matching against keys (i.e., $W^{K}x_{i}$) and $l_{S}^{C}$ is the list embedding generated by CPPM.

\begin{figure}[]
    \centering
     \setlength{\abovecaptionskip}{0.0cm}
  \setlength{\belowcaptionskip}{-0.6cm}
    \includegraphics[scale=0.35]{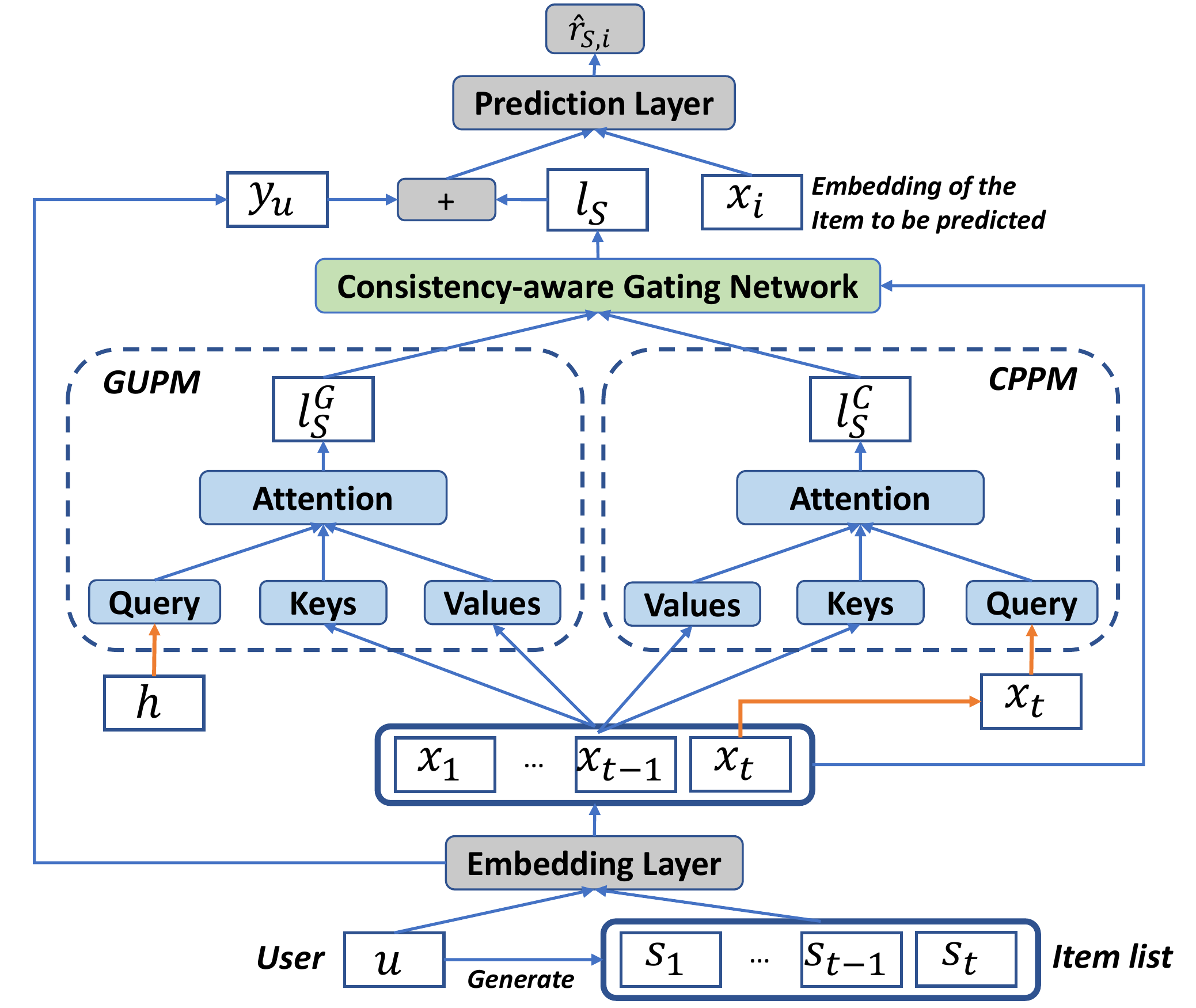}
    \caption{Framework of CAR for user-generated item list continuation.}
    \label{framework figure}
\end{figure}

\subsection{Consistency-aware Gating Network}
\label{Coherence-aware Gating Network}
As observed in Section \ref{motivation}, the list consistency varies greatly both within a single platform and across different platforms. Hence, the mixture of GUPM and CPPM may be better to handle this heterogeneity of list consistency. Since GUPM is designed for strongly consistent lists while CPPM is designed for weakly consistent lists, a natural way to intelligently fuse them is to consider the degree of the list consistency. Here, we propose a novel consistency-aware gating network for assigning different weights for the two models. The key question is how to feed the consistency into the gating network: 

\medskip
\noindent\textbf{Consistency-based Input:} We have applied the average of similarities between the last item and previous items for measuring the consistency in Eq.\eqref{consistency equation}, which can also be viewed as the similarity between the last item and the centroid of a list in Eq.\eqref{consistency in model equation idea}. The intuition here is that if the last item is different from the centroid of a list then the user preference has deviated from previous items. 
\begin{equation}
\label{consistency in model equation idea}
	\sum_{i=1}^{t}\frac{sim(x_{i}, x_{t})}{t} = sim(\sum_{i=1}^{t} \frac{x_{i}}{t}, x_{t})
\end{equation}

\noindent where $x_{t}$ is the embedding of the last item and $(\sum_{i=1}^{t} \frac{x_{i}}{t}) \in \mathbb{R}^{d}$ is the centroid embedding of a list. There are many ways to calculate the distance between two vectors like Manhattan distance or Euclidean distance. However, our goal here is not only to obtain the distance but also keeping the dimension of the vectors, which could be fed into the gating network for learning the weights. Hence, we use the difference between the two vectors to represent the consistency:

\begin{equation}
\label{consistency in model equation}
	z^{C}  = \sum_{i=1}^{t} \frac{x_{i}}{t} - x_{t}
\end{equation}
where $z^{C} \in \mathbb{R}^{d}$ is the consistency-based input.

\medskip
\noindent\textbf{List-embedding-based Input:} Items within a list also contain rich information to guide a recommender in fusing general and current preferences and should be fed into the gating network. The gating network receives a vector ($z \in \mathbb{R}^{d}$) as input so that it is inappropriate to pack item embeddings into a matrix as the input. Therefore, another attention network is firstly used to compress item embeddings into a list embedding:

\begin{equation}
\label{list embedding-based input equation}
	z^{L} = \sum_{i=1}^{t} softmax((W_{G}^{K}x_{i})^{T}h^{G})x_{i}
\end{equation}
where $z^{L} \in \mathbb{R}^{d}$ is the list embedding-based input,$W_{G}^{K} \in \mathbb{R}^{d \times d}$ is the projection to map $x_{i}$ into ``keys" space in this gating network and $h^{G} \in \mathbb{R}^{d}$ is the learnable vector to measure the weights of $x_{i}$. The final input is formed by concatenating $z^{C}$ and $z^{L}$ together: $z = z^{C} \oplus z^{L}$. Then, following \cite{jacobs1991adaptive, ma2018modeling}, the gating networks are linear transformations of the input with a softmax layer:
\begin{equation}
\label{gating network equation}
	g(z) = softmax(W^{G}z)
\end{equation}
where $g(z) = [g(z)_{C}, g(z)_{G}] \in \mathbb{R}^{2}$ is the vector of gating values, $W^{G} \in \mathbb{R}^{2 \times d}$ is the transformation matrix.

\subsection{CAR: Consistency-aware Attention-based Recommender}
\label{CAR section}
In Section \ref{Attention-based User Preference Model}, we propose two attention-based user preference models: the general preference model $l_{S}^{G}$ and the current preference priority model $l_{S}^{C}$. In this section, we adaptively fuse these two by weighted-summing together via the consistency-aware gating network in Section \ref{Coherence-aware Gating Network}:
\begin{equation}
	l_{S} = g(z)_{C} \times l_{S}^{C} + g(z)_{G} \times l_{S}^{G}
\end{equation}
where $l_{S}  \in \mathbb{R}^{d}$ is the final list embedding encoded with both general and current user preferences and $g(z) = [g(z)_{C}, g(z)_{G}]$ are gating values.


\medskip
\noindent\textbf{User Embedding:} So far, we only use item embeddings and list embeddings (aggregated from the item embeddings). Here, we add the user embedding of the list creator and list embedding together:
\begin{equation}
\label{user embedding equation}
	l_{S}^{u} = l_{S} + y_{u}
\end{equation}
where $y_{u}$ is the embedding of the user who generates list $S$. The intuition is that adding the user embedding may introduce more personalization information and provide better recommendation performance. Whether using this user embedding depends on the performance at the validation set and the impact of it will be discussed in Section \ref{ablation study section}.

\medskip
\noindent\textbf{Prediction Layer and Loss Function:} For CUPM, CPPM and CAR, we apply the same prediction layer and loss function for fair comparison. List embedding $l_{S}$ is firstly fed into a two-layer feed-forward network to introduce nonlinearity into the model, following \cite{vaswani2017attention, kang2018self}. We have:
\begin{equation}
\label{feed-forward equation}
	f^{u}_{S} = ReLU(l_{S}^{u}W^{1} + b^{1})W^{2} + b^{2}
\end{equation}
where $W^{1}, W^{2} \in \mathbb{R}^{d \times d}$ and $b^{1}, b^{2} \in \mathbb{R}^{d}$ are parameters of the feed-forward network. After that, prediction score for item $i$ is calculated by a matrix factorization layer: $r_{S, i} = x_{i}^{T}f^{u}_{S}$. Finally, the classic pairwise ranking loss used in BPR \cite{bpr} is applied as our objective function:
\begin{equation}
	-\sum_{S\in S^{all}}\sum_{t=1}^{N} log(\sigma(r_{S, i} - r_{S, j}))
\end{equation}
where $S^{all}$ denotes the set of all lists, $r_{S, j}=x_{j}^{T}f^{u}_{S}$ and $x_{j}$ is the embedding of randomly sampled negative item $j$ and $\sigma$ is the sigmoid function.

\medskip
\noindent\textbf{Model Training}
Adam \cite{kingma2014adam} is applied as the optimizer. We perform mini-batch training, where each batch contains a group of lists. To generate more training samples, for each list, starting from the second item, each item is treated as the test item ($s_{t+1}$) and items before this item are treated as the training data. To control the length of lists in mini-batch, the most recent $n$ items are considered. If the list length is less than $n$, a `padding' item is repeatedly added until the length is $n$.

\section{Experimental Setup}
In this section, we introduce the experimental setup, including datasets, metrics, baselines, and reproducibility details.  


\subsection{Datasets}
\label{datasets}



We first select one relatively new dataset from the literature of automatic playlist generation and continuation:

\begin{itemize}
	\item \textbf{Art of the Mix} \footnote{http://www.artofthemix.org/} is a website where users can upload their song playlists. \textbf{AOTM}\footnote{https://bmcfee.github.io/data/aotm2011.html} dataset is proposed by \cite{mcfee2012hypergraph} in 2011, collecting playlists from Jan 1998 to June 2011, resulting in totally 101,343 unique playlists. The main reason for choosing this dataset is that each list was generated by a user not a commercial radio DJ, which is an accurate sample of real playlists that occur in our daily life \cite{mcfee2011natural}. (Note that datasets used in the RecSys Challenge 2018 \cite{chen2018recsys} have not been released.)
\end{itemize}


We also crawl and build three datasets, which are available at here\footnote{https://github.com/heyunh2015/ListContinuation\_WSDM2020}. These datasets are drawn from three popular platforms with different kinds of lists (book-based, song-based, and answer-based), reflecting a variety of scenarios to evaluate our approach.

\begin{itemize}
	\item \textbf{Goodreads} is a popular site for book reviews and recommendation where users can curate correlated books into booklists. 
	\item \textbf{Spotify} is a music streaming service where users can create playlists and curate songs into them. 
	\item \textbf{Zhihu} is 	a question and answer community where users can select correlated answers to form an answer list. For example, a list named ``Food" may contain answers as the response to two questions correspondingly: ``How to cook steak?" and ``Who is the barbecue king of Texas?". 
\end{itemize}

Crawling all data generated in a short time window (e.g., one or two weeks) is widely used to sample a moderate-size  dataset for research from millions or even billions of data \cite{zhong2013sharing}. However, since the generation of a list could take its owner months or years to occasionally curate hundreds of items, crawling all lists created in a short time window  would obtain relatively short ``immature" lists. We first randomly sample a batch of users and then crawl lists either created or followed by them.

For AOTM-2011, Goodreads and Zhihu datasets, following \cite{kang2018self}, we filter out items appearing fewer than 5 times and lists shorter than 5 items. For Spotify, the largest dataset, we filter out items appearing fewer than 20 times and lists shorter than 20 items. For extreme long lists, we truncate them by keeping the first 1,000 items. Table \ref{Statistics of datasets} summarizes these datasets, where any one of them has a million-size list-item interactions.

\begin{table}[htbp]
  \centering
	 \setlength{\abovecaptionskip}{0.0cm}
  \setlength{\belowcaptionskip}{-0.30cm}
	\small
	\renewcommand\arraystretch{0.80}
	\setlength{\tabcolsep}{2.0pt}
  \caption{Statistics of datasets}
    \begin{tabular}{lcccc}
	\toprule
	& \multicolumn{1}{c}{ AOTM } & \multicolumn{1}{c}{Goodreads} & \multicolumn{1}{c}{Spotify} & \multicolumn{1}{c}{Zhihu} \\
	\midrule
    \#Users &              14,115  &                     12,525  &                 14,911  &                           9,561  \\
    \#Lists &              81,798  &                     21,877  &                 70,485  &                         31,040  \\
    \#Items &              64,181  &                   148,576  &               104,910  &                       119,147  \\
    \#List-item interactions &         1,030,596  &                2,441,308  &            7,287,212  &                    1,593,865  \\
    Avg. \# of lists per user &                    5.8  &                           1.7  &                       4.7  &                               3.2  \\
    Avg. \# items per list &                  12.6  &                       111.6  &                   103.4  &                             51.3  \\
    Density & 0.020\% & 0.075\% & 0.099\% & 0.043\% \\
	\bottomrule
	
    \end{tabular}%
  \label{Statistics of datasets}%
\end{table}%

\noindent\textbf{Data Partition:} We split each list into three parts: (1) the last item for testing, (2) the item right before the last one for validation, and (3) the rest of items for training. Note that during testing, the input sequences contain the training items and the validation item.

\begin{table*}[htbp]
\def\sym#1{\ifmmode^{#1}\else\(^{#1}\)\fi}
  \centering
	\small
	\setlength{\abovecaptionskip}{0.0cm}
  \setlength{\belowcaptionskip}{-0.30cm}
	\renewcommand{\arraystretch}{0.85}
	\setlength{\tabcolsep}{1.6pt}
  \caption{Performance of CAR and the baselines}
    \begin{tabular}{|c|cccc|cccc|cccc|cccc|}
	\toprule
          & \multicolumn{4}{c|}{AOTM}      & \multicolumn{4}{c|}{Goodreads} & \multicolumn{4}{c|}{Spotify}   & \multicolumn{4}{c|}{Zhihu} \\
	\midrule
    Metrics & HR@5  & N@5 & HR@10 & N@10 & HR@5  & N@5 & HR@10 & N@10 & HR@5  & N@5 & HR@10 & N@10 & HR@5  & N@5 & HR@10 & N@10 \\
	\midrule
    MostPop & 0.2632 & 0.1817 & 0.3806 & 0.2194 & 0.2379 & 0.1666 & 0.3308 & 0.1964 & 0.2884 & 0.1983 & 0.4060 & 0.2362 & 0.4061 & 0.3003 & 0.5214 & 0.3375 \\
    MF    & 0.3821 & 0.2767 & 0.5002 & 0.3149 & 0.5900 & 0.4506 & 0.7105 & 0.4897 & 0.6432 & 0.4885 & 0.7686 & 0.5292 & 0.6135 & 0.5008 & 0.7059 & 0.5307 \\
    BPR   & 0.4711 & 0.3528 & 0.5965 & 0.3934 & 0.7229 & 0.5860 & 0.8207 & 0.6178 & 0.7367 & 0.5788 & 0.8495 & 0.6155 & 0.6209 & 0.5014 & 0.7255 & 0.5352 \\
    NeuMF   & 0.4519 & 0.3338 & 0.5733 & 0.3730 & 0.6779 & 0.5235 & 0.7906 & 0.5602 & 0.5950 & 0.4375 & 0.7362 & 0.4833 & 0.6363 & 0.5078 & 0.7434 & 0.5425 \\
    CMN   & 0.3885 & 0.2852 & 0.5036 & 0.3223 & 0.7262 & 0.5738 & 0.8277 & 0.6069 & 0.7230 & 0.5629 & 0.8385 & 0.6005 & 0.6633 & 0.5323 & 0.7653 & 0.5655 \\
    transRec & 0.3704 & 0.2664 & 0.4944 & 0.3064 & 0.3287 & 0.2374 & 0.4398 & 0.2731 & 0.4012 & 0.2922 & 0.5224 & 0.3314 & 0.5033 & 0.3856 & 0.6200 & 0.4234 \\
    GRU4Rec & 0.3013 & 0.2162 & 0.4091 & 0.2509 & 0.5131 & 0.4187 & 0.6140 & 0.4513 & 0.7584 & 0.6458 & 0.8403 & 0.6724 & 0.5603 & 0.4498 & 0.6647 & 0.4836 \\
    Caser & 0.3678 & 0.2615 & 0.4980 & 0.3035 & 0.6118 & 0.4659 & 0.7442 & 0.5090 & 0.7094 & 0.5598 & 0.8193 & 0.5955 & 0.6461 & 0.5042 & 0.7709 & 0.5447 \\
    SASRec & 0.4977 & 0.3650 & 0.6298 & 0.4077 & 0.7808 & 0.6449 & 0.8599 & 0.6707 & 0.8019 & 0.6577 & 0.8904 & 0.6865 & 0.7355 & 0.5899 & 0.8331 & 0.6217 \\
	\midrule
    CAR   & 0.5505\sym{*} & 0.4140\sym{*} & 0.6807\sym{*} & 0.4562\sym{*} & 0.7991\sym{*} & 0.6778\sym{*} & 0.8639 & 0.6990\sym{*} & 0.8113\sym{*} & 0.6758\sym{*} & 0.8945\sym{\circ} & 0.7029\sym{*} & 0.7468\sym{\circ} & 0.6096\sym{*} & 0.8410 & 0.6402\sym{*} \\
	\bottomrule
	\multicolumn{5}{l}{\footnotesize \sym{\wedge} \(p<0.05\), \sym{\circ} \(p<0.01\), \sym{*} \(p<0.001\)}\\
    \end{tabular}%
  \label{all results}%
\end{table*}%

\subsection{Evaluation Protocol}
In real world like YouTube recommendation \cite{covington2016deep}, it is common to only calculate prediction scores for items from a candidate set and rank them for the final recommendation, since it will be impractical to rank the whole millions of items in real world. The candidate generation model is not the emphasis of this work, hence we follow \cite{kang2018self, ncf} to randomly sample 100 negative items and rank these items with the ground-truth item.

\noindent\textbf{Metrics:} For each list, each tested algorithm selects K items ordered by prediction scores from the candidate items and then match the top K items against the ground-truth item. We adopt two evaluation metrics: Normalized Discounted Cumulative Gain (NDCG) at 5 (N@5) and 10 (N@10), and Hit Rate at 5 (HR@5) and 10 (HR@10).

\subsection{Baseline}
In this section, we introduce a suite of baselines. Except to traditional item-based recommendation methods include \textbf{ItemPop} (most popularity), \textbf{MF} \cite{koren2009matrix} and \textbf{BPR} \cite{bpr}, we also have:

%
%

\noindent\textbf{NeuMF.} Neural collaborative filtering (NeuMF) \cite{ncf} concatenates latent factors learned from a generalized matrix factorization model and a multi-layered perceptron model.

\noindent\textbf{NMN.} Neural Memory Network (NMN) \cite{ebesu2018collaborative} is a state-of-the-art memory-based neural network to unify the global latent factor model and local neighborhood structure.

\noindent\textbf{TransRec.} Translation-based Recommendation (TransRec) \cite{he2017translation} models the third order interactions between a list, the item previously curated and the next item to curate.

\noindent\textbf{GRU4Rec.} GRU4Rec \cite{hidasi2015session} is a GRU-based \cite{cho2014properties} RNN model for session-based recommendation (predict the last item for testing sessions). We treat each list as a session.

\noindent\textbf{Caser.} Convolutional Sequence Embeddings (Caser) \cite{tang2018personalized} is a CNN-based model for sequential recommendation, which captures high-order markov chains by convolutional operations applied on the most recent items.

\noindent\textbf{SASRec.} Self-Attention based Sequential Recommendation model (SASRec) apply self-attention mechanism \cite{vaswani2017attention} for next item recommendation and achieves state-of-the-art performance, with large improvements over GRU4Rec and Caser observed in their paper.

\noindent\textbf{Other Related Work but with Different Settings:} Note that we are aware of the recent progress on automatic playlists continuation in \cite{tran2019adversarial}. However, their goal is to recommend a song to a list where order information is absent (i.e., a collection of songs). Since this goal is quite different from our continuing ordered items, we do not compare with it. Besides, many traditional playlists generation work (e.g., \cite{chen2012playlist, mcfee2011natural, mcfee2012hypergraph}) focus on estimating the likelihood of a new playlist, which is also different and not compared.

\subsection{Reproducibility}
\textit{All datasets and code of our approach can be found at here\footnote{https://github.com/heyunh2015/ListContinuation\_WSDM2020}.} We implement CAR with Tensorflow. We implement ItemPop, BPR, and MF. The code of NeuMF\footnote{https://github.com/hexiangnan/neural\_collaborative\_filtering}, NMN\footnote{https://github.com/tebesu/CollaborativeMemoryNetwork}, Caser\footnote{https://github.com/graytowne/caser\_pytorch} and SASRec\footnote{https://github.com/kang205/SASRec} are from the authors. For TransRec\footnote{https://github.com/YifanZhou95/Translation-based-Recommendation} and GRU4Rec\footnote{https://github.com/Songweiping/GRU4Rec\_TensorFlow}, we also use code from public resources. All neural network models were trained using Nvidia GeForce GTX Titan X GPU with 12 GB memory.


\medskip
\noindent\textbf{Parameter Settings.} For CAR and baseline methods, all hyper-parameters are tuned on the validation dataset, where early stopping strategy is applied such that we terminate training if validation performance does not improve over 10 iterations. For CAR: the batch size is tested from \{16, 32, 64, 128, 256\} and 128 is selected for all datasets according to the results on the validation dataset. The learning rate is 0.001 for all datasets. Following \cite{kang2018self}, the candidates for the latent dimensionality $d$ is from \{10, 20, 30, 40, 50\} and we select 50 for all datasets. The maximum sequence length $n$ for all datasets is set to 500. For NeuMF, the number of MLP layers is set to 3. For CMN, the number of hops is set to 3. The Markov order of GRU4Rec and Caser is selected from \{1,\text{...}, 9\}. For SASRec, the maximum length of handled sequence is set to 500 and the number of blocks is set to 2. 


\section{Experimental Results and Analysis}

In this section, we present our experimental results and discussion toward answering the following research questions (RQs):
\begin{itemize}
	\item \textbf{RQ1:} How well does CAR perform for continuing user-generated item lists compared to baseline methods?
	\item \textbf{RQ2:} What is the impact of the design choices of CAR on the quality of list continuation? For example, is the consistency-aware gating network effective for fusing the two models?
	\item \textbf{RQ3:} What is the impact of list consistency? Specifically, for lists with the weak consistency, is the current preference priority model superior to the general preference model (and vice versa)?
	\item \textbf{RQ4:} What is the impact of the hyper-parameters of CAR on the quality of list continuation.
\end{itemize}


\subsection{RQ1: List Continuation Performance}
Experimental results of CAR and baseline methods are presented in Table \ref{all results}. The results show that CAR outperforms all baseline. For example, CAR achieves an NDCG@5 of 0.4140 compared to the second best performance 0.3650 from SASRec. The results of a two-sided significant test between CAR and the strongest baseline SASRec is also presented, where \text{$\wedge$} means the p-value is smaller than 0.05, \text{$\circ$} means the p-value is smaller than 0.01 and \text{$*$} means the p-value is smaller than 0.001. We see that the CAR approach significantly outperforms the strongest baselines in nearly all metrics. We observe that SASRec is the most competitive method with CAR, which is not surprising because SASRec is also an attention-based model that, to a certain extent, overcomes the long-term dependency problem suffered by CNN and RNN based models \cite{vaswani2017attention}. However, SASRec is proposed for sequential user action (e.g., purchase) prediction where current user preference (last purchased item) is often more important than general preference. Hence, for lists with strong consistency among the items (e.g., playlist ``The lion king music"), SASRec may not fully utilize the user preference information from previous items. To sum up, the main improvement of CAR compared with SASRec is that we intelligently fuse current and general user preference together and further enhanced by the consistency between recent items and previous items.

\subsection{RQ2: Ablation Study}
\label{ablation study section}
\begin{table}[htbp]
  \centering
   \setlength{\abovecaptionskip}{0.0 cm}
  \setlength{\belowcaptionskip}{-0.35cm}
  \renewcommand\arraystretch{0.80}
\small
\setlength{\tabcolsep}{2.0pt}
  \caption{Ablation Study}
    \begin{tabular}{ccccc}
	\toprule
          & AOTM  & Goodreads & Spotify & Zhihu \\
	\midrule
    CAR   & 0.4140 & 0.6778 & 0.6758 & 0.6096 \\
    No Gating Network & 0.3960 & 0.6734 & 0.6688 & 0.6035 \\
    CPPM & 0.3935 & 0.6578 & 0.6448 & 0.5896 \\
    GUPM  & 0.3921 & 0.6524 & 0.6446 & 0.5812 \\
	\bottomrule
    \end{tabular}%
  \label{ablation study table}%
\end{table}

Table \ref{ablation study table} presents the results in terms of NDCG@5 of the ablation study at CAR. \textbf{Mixture of the Models:} We firstly compare GUPM (general preference model) and CPPM (current preference priority model) against ``No Gating Network" model, which simply adds CPPM and GUPM together without using the consistency-aware gating network (i.e., $l_{S} = l_{S}^{C} + l_{S}^{G} $). We observe that the simple mixture of the two models outperforms any one of them. This shows that CPPM and GUPM do complement each other to provide a better overall performance. 

\medskip
\noindent\textbf{Consistency-aware Gating Network:} Then, we observe that CAR is superior to the version without the gating network. For example, CAR obtains 0.4140 in terms of NDCG@5 which is better than 0.3960 obtained by ``No Gating Network". This comparison shows that our proposed consistency-aware gating network is effective to intelligently fuse the current and general user preferences. 

\begin{figure}[]
    \centering
    \setlength{\abovecaptionskip}{0.01cm}
  \setlength{\belowcaptionskip}{-0.15cm}
    \includegraphics[width=3.05in]{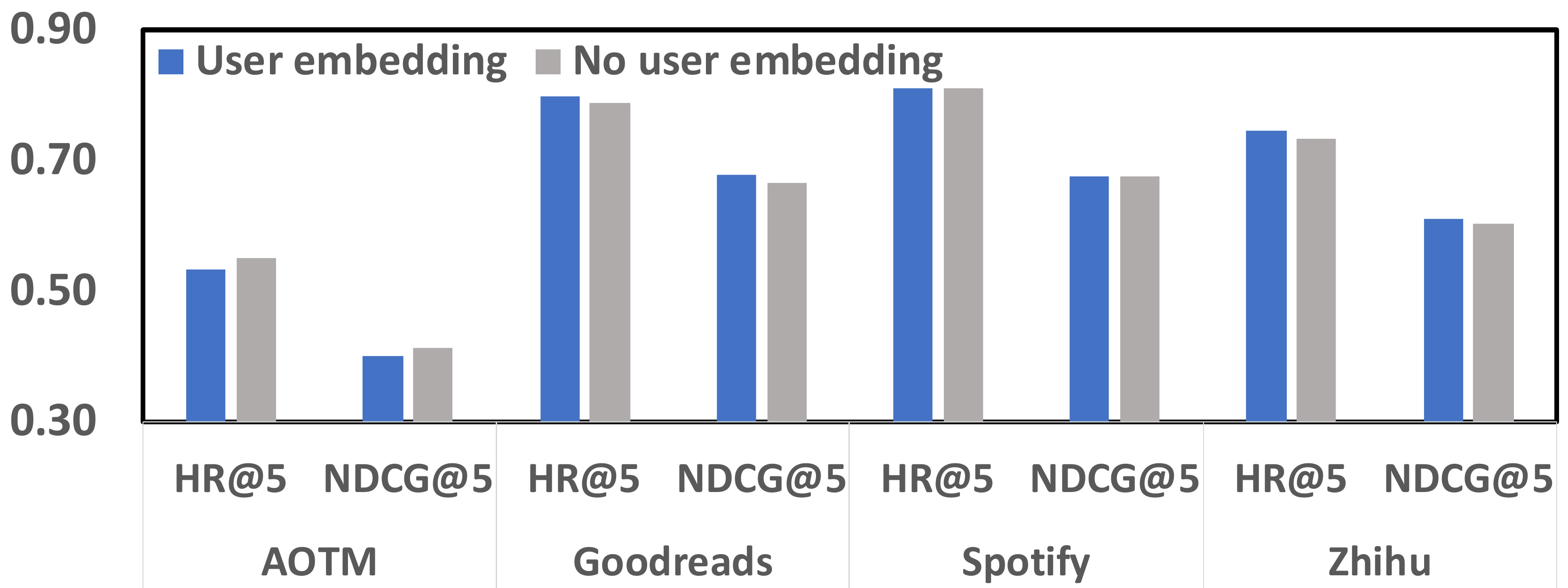}
    \caption{Impact of User Embedding}
    \label{user embedding impact figure}
\end{figure}


\medskip
\noindent\textbf{User Embedding:} We study the impact of user embedding introduced in Eq.\eqref{user embedding equation}. Users have different preferences for generating item lists. Hence, introducing user embeddings into the model of user preferences at each list is supposed to improve the personalization and provide better performance. However, in Figure \ref{user embedding impact figure}, we observe that the user embedding harms the performance for Goodreads and Zhihu, achieves the same performance for Spotify and only improves the performance for AOTM. Table \ref{Statistics of datasets} shows that each user create 5.8 lists at AOTM, 4.7 lists at Spotify, 3.2 lists at Zhihu, and 1.7 lists at Goodreads. Hence, we observe that the more lists created by each user, the better performance can be obtained by adding the user embeddings. This is reasonable because the more lists a user creates the more personalization information of the user can be learned.



\subsection{RQ3: Impact of List Consistency}
In this section, we design an experiment to evaluate our intuition about the impact of list consistency on the quality of predicting the next item, presented in the introduction, that for lists with weak consistency between recent items and previous items, the user preference has changed and thus the recent items should be paid more attention. For these weakly consistent lists, we propose a current preference priority model (CPPM). And for lists with strong consistency, all of the curated items could be treated equally no matter curated early or late. Correspondingly, we propose general preference model (GUPM) for these strongly consistent lists. Hence, a natural way to evaluate the intuition is to compare the consistency between the lists where GUPM wins and the lists where CPPM wins.

\begin{figure}
  \centering
   \setlength{\abovecaptionskip}{0.01cm}
  \setlength{\belowcaptionskip}{-0.15cm}
 \subfigure[AOTM]{
    \label{AOTM Case}
    \includegraphics[width=1.5in]{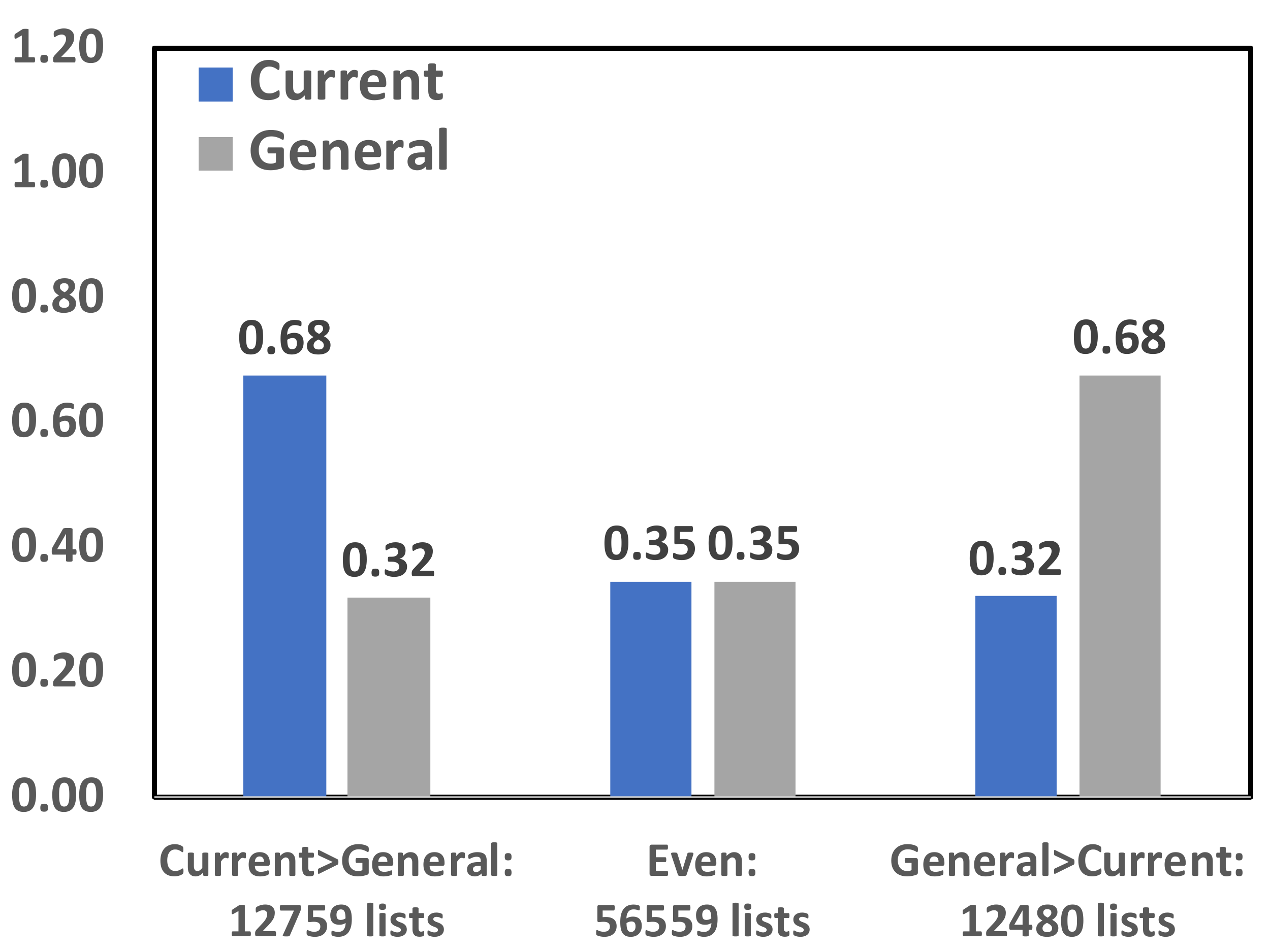}}
  \subfigure[Goodreads]{
    \label{Goodreads Case} 
    \includegraphics[width=1.5in]{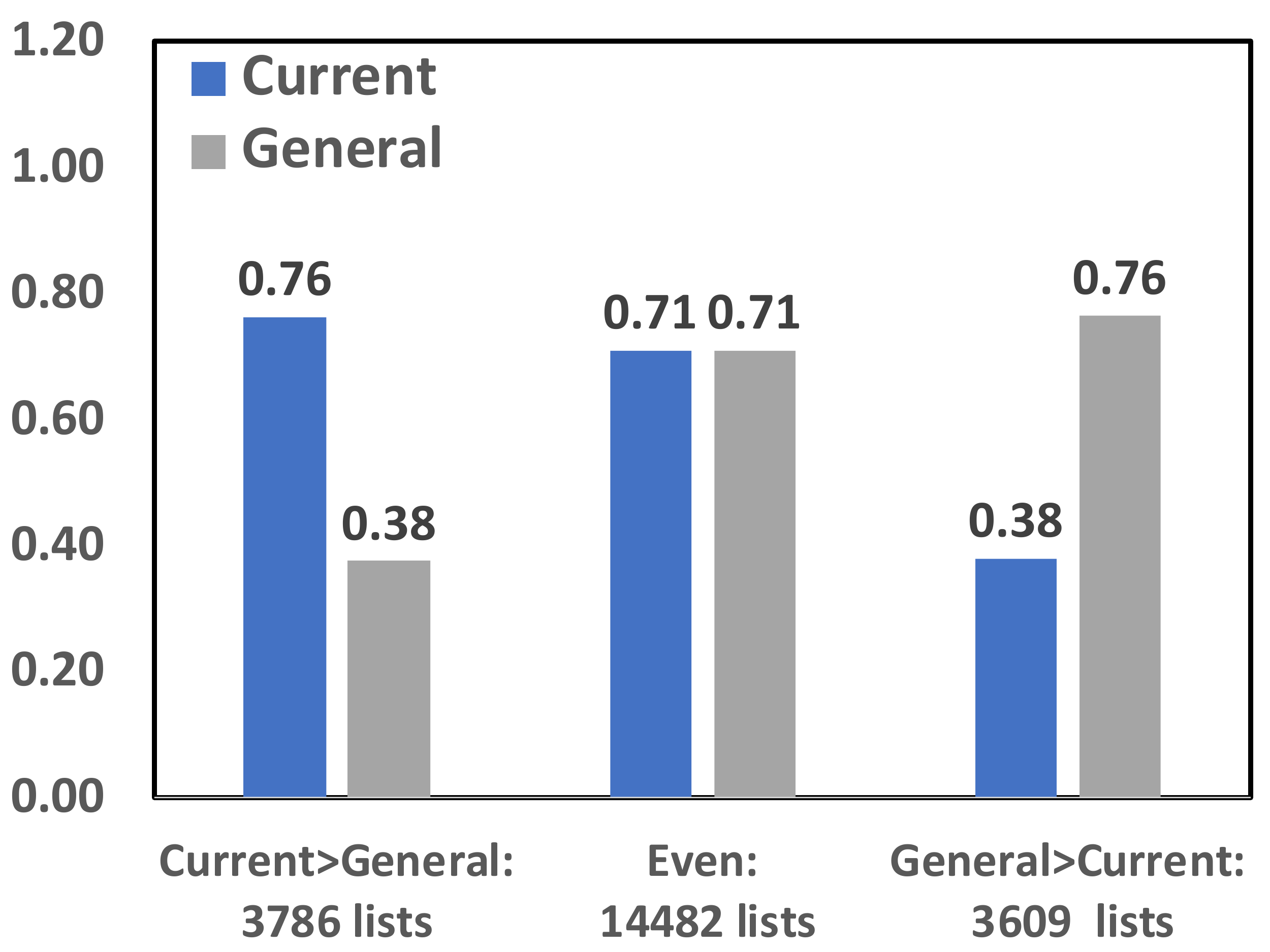}}
   \caption{Performance in terms of NDCG@5 at cases that GUPM wins (General>Current), even and CPPM wins (Current>General). Each model has a large advantage over the other one on a bunch of lists. Same observation is obtained at Spotify and Zhihu.}
  \label{Case Performance} 
\end{figure}

\medskip
\noindent\textbf{Case by Case Comparison:} We firstly compare the performance (taking NDCG@5 as an example) of GUPM and CPPM at each list to find out the lists where GUPM wins and the lists where CPPM wins. In Figure \ref{Case Performance}, we observe that the two models both have a large advantage over the other one on a section of lists. Taking AOTM dataset as an example, GUPM is superior to CPPM with a large margin of 0.68 against 0.32 on a different set of 12,480 lists, while CPPM achieves 0.68, largely outperforming 0.32 obtained by GUPM on 12,759 lists. Such a huge divergence of the performance suggests that the two sets of lists are different from each other.

\begin{figure}[]
     \centering
     \setlength{\abovecaptionskip}{0.01cm}
   \setlength{\belowcaptionskip}{-0.15cm}
     \includegraphics[width=3.05in]{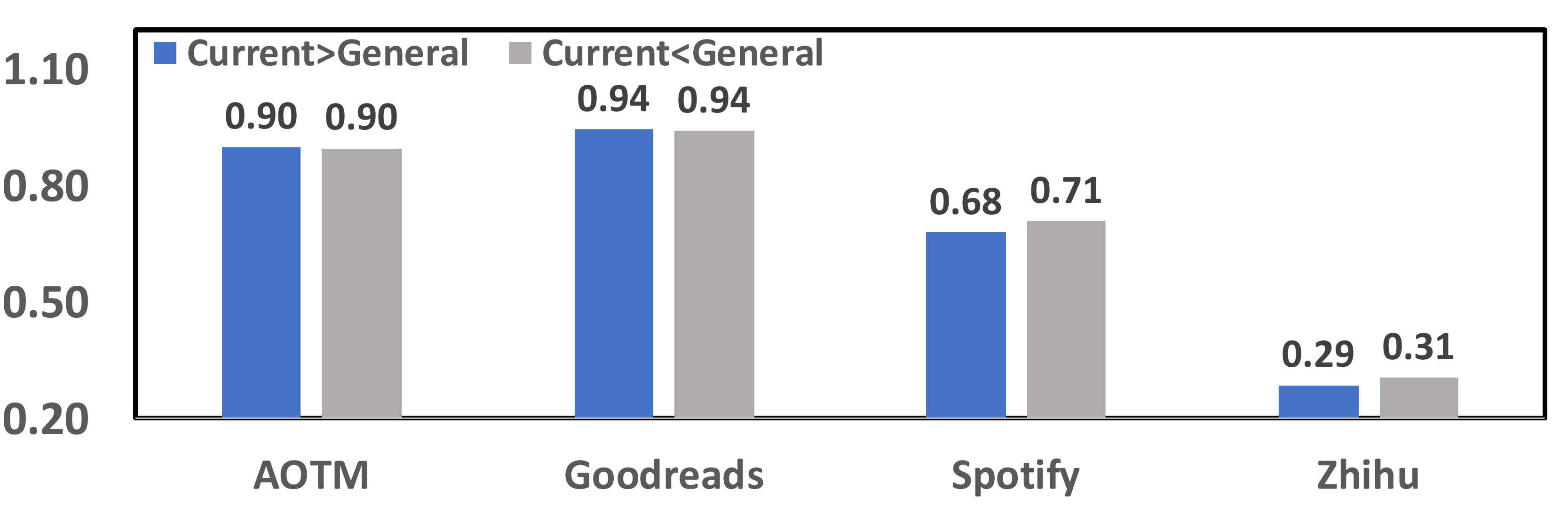}
     \caption{Comparison of average consistency between the cases that CPPM wins (Current>General) and the cases that GUPM wins (General>Current).}
     \label{consistency comparison figure}
 \end{figure}

\medskip
\noindent\textbf{Consistency Comparison:} Hence, we calculate the consistency between the last item and previous items according to Eq.\eqref{consistency equation} and compare the average consistency of the two bunches of lists in Figure \ref{consistency comparison figure}. Note that we use the item embeddings from trained word2vec model in Section \ref{motivation}, rather than embeddings from our approach, in order to provide a more objective perspective for measuring the consistency. Interestingly, in AOTM and Goodreads with commonly strong consistency, we observe that the lists where GUPM wins have a similar consistency with the lists where CPPM wins. This is understandable that user general and current preference should generate similar results if a list is always consistent. However, in Spotify and Zhihu with commonly weak consistency, we observe that the lists where GUPM wins have a stronger consistency than the lists where CPPM wins. Taking Spotify as an example, the lists where GUPM wins have an average consistency of 0.71 while the lists where CPPM wins have a weaker consistency of 0.68. Hence, the comparison between the average consistency of GUPM and CPPM in these two datasets supports our intuition of the impact of list consistency on the quality of predicting the next item.


\subsection{RQ4: Impact of Hyper-parameters}
\begin{figure}
  \centering
   \setlength{\abovecaptionskip}{0.01cm}
  \setlength{\belowcaptionskip}{-0.35cm}
  \subfigure[Dimension $d$]{
    \label{dimension} 
    \includegraphics[width=1.6in]{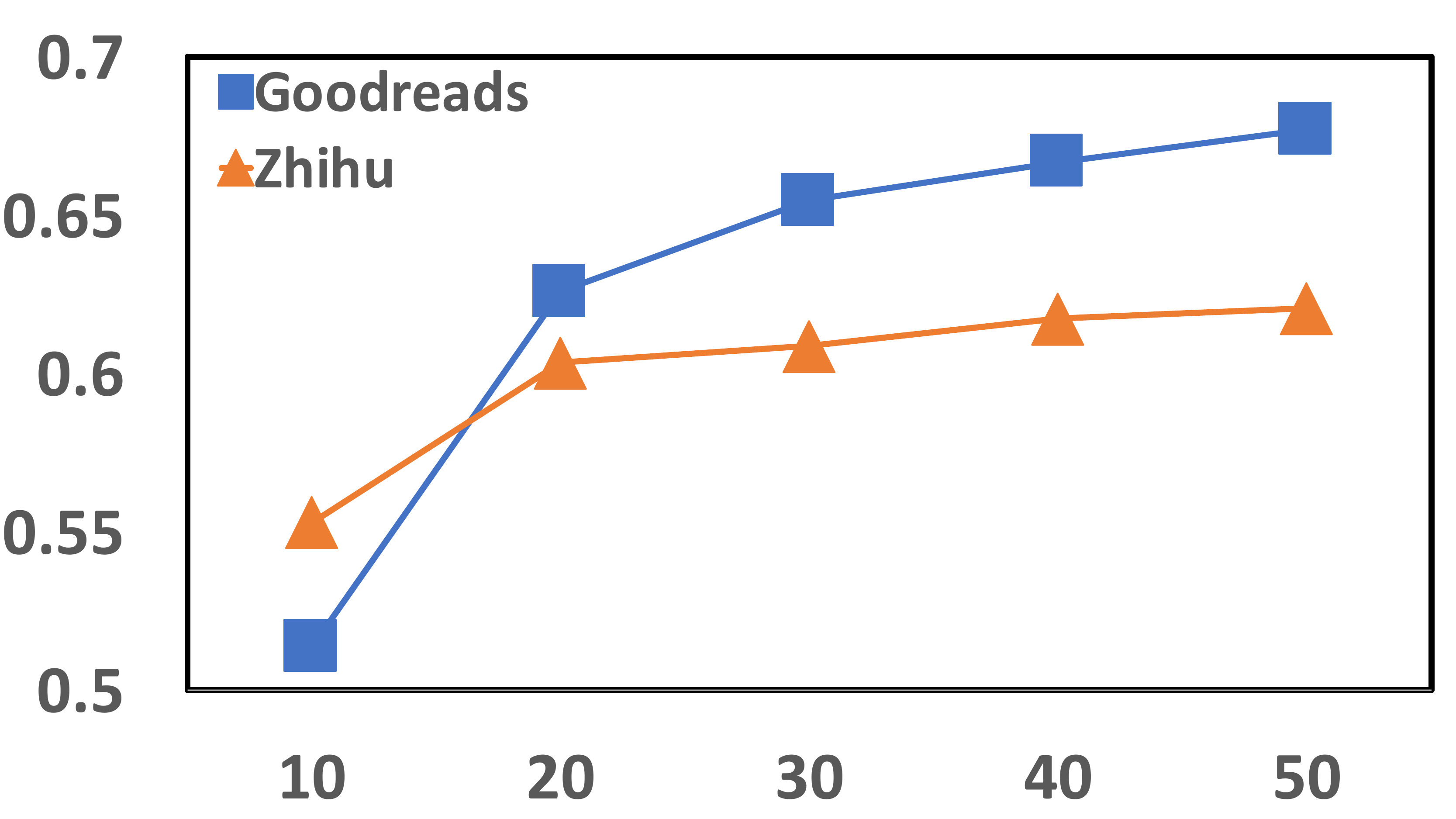}}
  \subfigure[Maximum Length of Item Sequence $n$]{
    \label{max length} 
    \includegraphics[width=1.6in]{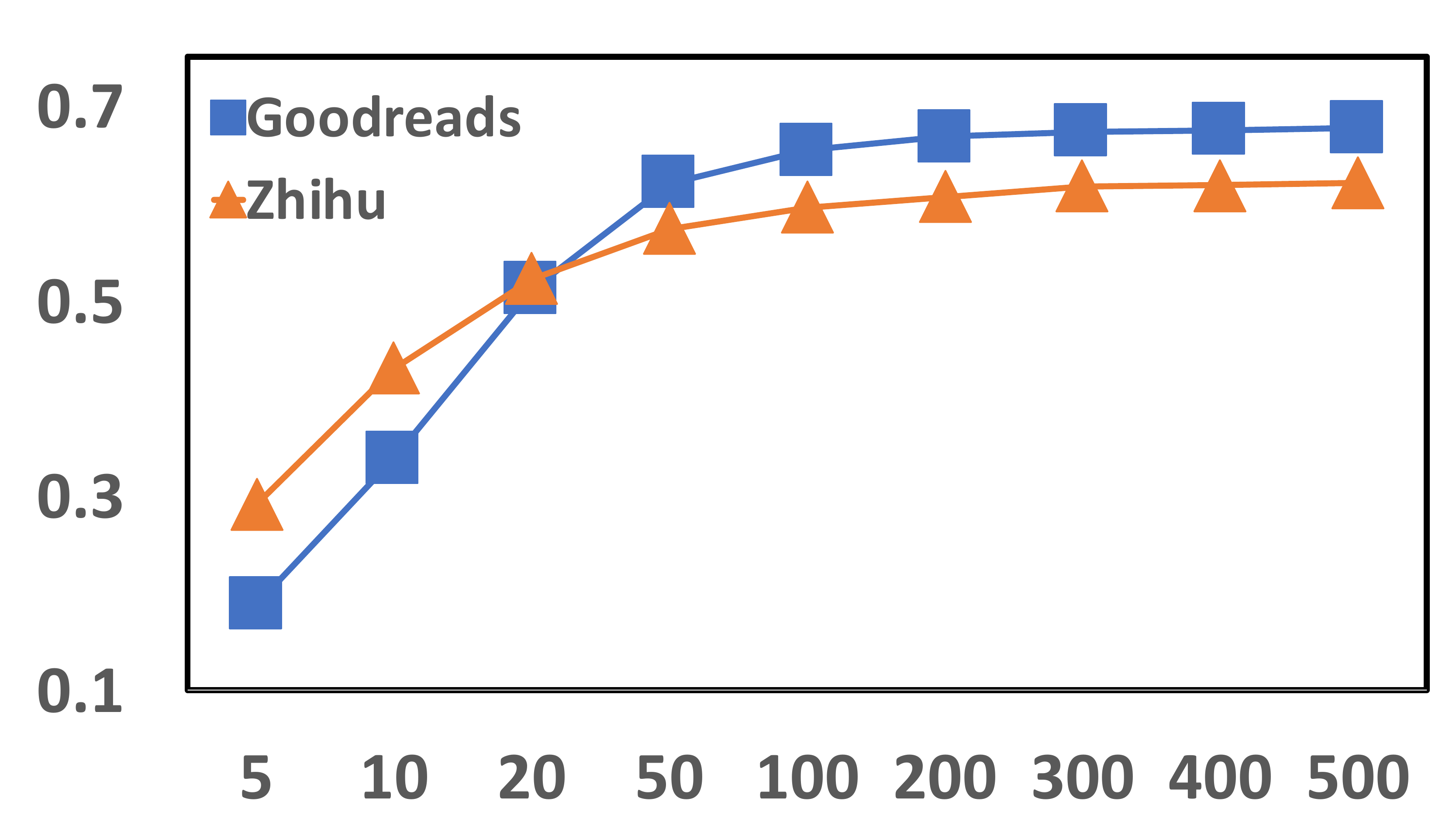}}
   \caption{Impact of hyper-parameters in terms of NDCG@5 taking Goodreads and Zhihu as examples.}
  \label{Impact of hyper-parameters} 
\end{figure}

In this section, we analyze the impact of hyper-parameters. Due to the limited space, two important hyper-parameters are selected to discuss. In Figure \ref{dimension}, we observe that our approach benefits from larger numbers of the latent dimensionality $d$. The best results are obtained with $d=50$ in Goodreads and Zhihu. In Figure \ref{max length}, we observe that the longer the item sequence that is fed into our model, the better performance can be obtained for predicting the next item. The best results are achieved with $n=500$ in Goodreads and Zhihu and the performance decreases largely when $n<100$. This demonstrates that user's general preference always maintains influence on curating items in a list and items that are curated even a long time ago should also be paid attention for list continuation.

\section{Conclusion and Future Work}
In this paper, we focus on automatically continuing user-generated item lists, i.e., predicting the next possible item for each list. We observe that the consistency between recent and previous items reflects the dynamic change of user preferences. Specifically, if recent items are not similar with previous items, the user preference has probably changed and hence we should pay more attention on the recent items for predicting the next item. Therefore, we are motivated to propose a consistency-aware attention-based recommender (CAR), where a novel consistency-aware gating network is designed for capture the discontinuity between the current user preference and the general user preference. The evaluation on four datasets demonstrate the effectiveness of CAR versus state-of-the-art alternatives. In the future, we will utilize some side information to enhance CAR, like friend relationships among users and text-based titles of user-generated item lists.